\documentstyle[twoside,fleqn,epsf,espcrc2]{article}

\def\be		{\begin{equation} }
\def\ee		{\end{equation} }
\def\bea		{\begin{eqnarray} }
\def\eea		{\end{eqnarray} }
\def\indexsize	{\tiny}
\def\eff		{\mbox{\indexsize eff} }

\def\dof		{d.o.f. }

\newlength{\figsize}
\newlength{\figoffset}
\newlength{\figbackup}
\newlength{\figendsp}

\def\preprints{
\vspace{-16ex}
{\small
\begin{tabbing}
\` {\sl hep-lat/9808050} \\
    \\
\` LSUHE No. 273--1998 \\
\` OUTP--98--65P \\
\` August 1998 \\
\end{tabbing} 
}
\vspace*{0.1in}
}

\title{
\preprints
The string tension in the maximally Abelian gauge after smoothing.}

\author{\underline{A. Hart}%
	\address{Dept. of Physics and Astronomy, 
	Louisiana State University, 
      Baton Rouge, LA 70803, USA.},
      J.D. Stack%
	\address{Dept. of Physics, University of Illinois 
	at Urbana--Champaign, 1110 W. Green Street, 
      Urbana, IL 61801, USA.\\}
	and 
      M. Teper%
	\address{Theoretical Physics, University of Oxford, 
	1 Keble Road,
	Oxford OX1 3NP, UK.\\}.}

\begin{document}

\begin{abstract}
\noindent
We apply smoothing to SU(2) lattice field configurations in 3+1 
dimensions before fixing to the maximally Abelian gauge. The Abelian 
projected string tension is shown to be stable under this, whilst the 
monopole string tension declines by ${\cal O}(30\%)$. Blocking of the 
SU(2) fields reduces this effect, but the use of extended monopole 
definitions does not. We discuss these results in the context of 
additional confining excitations in the U(1) vacuum.
\end{abstract}

\maketitle

In recent years a confinement mechanism based on a dual superconducting
vacuum 
\cite{mandelstam76,thooft81}
has been studied, motivated in large part by the observation 
that in SU(2)in the maximally Abelian (MA) gauge
\cite{kronfeld87}, 
degrees of freedom (d.o.f.)
may be numerically integrated out (`Abelian projection') to
yield U(1) configurations reproducing the original string
tension, and that magnetic monopoles 
are responsible for this
\cite{suzuki90,stack94,bali96}.
A physically attractive scenario stems from the assumption
that the non--Abelian theory may be written as a combination of 
UV and IR \dof 
All the IR physics is postulated to remain in the residual 
U(1) fields after fixing to the MA gauge. The full SU(2) 
string tension should then be calculable from an Abelian Wilson 
loop in this gauge.

Application of short range perturbations to the SU(2) fields cannot 
alter the asymptotic string tension, nor is it expected to change 
any estimate of this from a correlation function extending over a scale
much greater than the range of the perturbation. A crucial test of the
above scenario is that the U(1) string tension after gauge fixing behaves
similarly, and that Abelian dominance remains.

Smoothing of the SU(2) fields is such a perturbation.
We generate an ensemble of SU(2) field configurations using
the Wilson plaquette action at $\beta = 2.5$ on $16^4$, $20^4$; 
lattices large enough that we may estimate the asymptotic string 
tension.
To this ensemble is applied one Wilson action smoothing (= cooling) 
step.
Each link in (`staggered') turn on the lattice is updated to locally 
minimise the Wilson action. This is a local smoothing on the scale of 
one lattice spacing and will not affect the string tension which we 
extract on a scale of several lattice spacings
\cite{teper94}. 

In this work smoothed configurations are fixed to the MA gauge 
and Abelian projected to create an ensemble of U(1) fields and
the magnetic monopole worldlines are identified as in
\cite{hart98b}.
We define effective U(1) and monopole string tensions from the 
square Creutz ratios:
$
K_{\eff}(r) = - \ln C(r,r),
$
which in the limit of large $r$ converge to the 
asymptotic string tensions, $K = a^2 \sigma$. 

\begin{figure}[t]

\leavevmode
\begin{center}

\hbox{%
\hspace{\figoffset}
\epsfxsize = \figsize
\epsffile{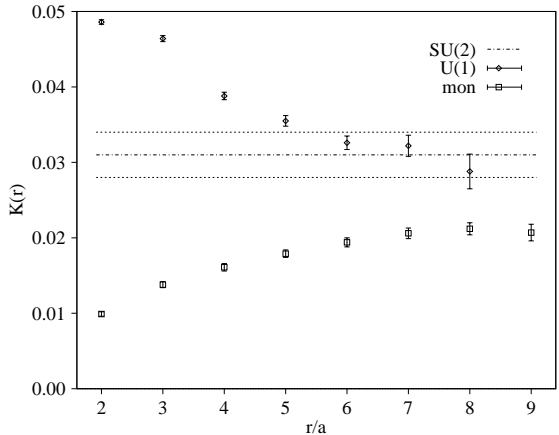}
}

\vspace{\figbackup}
\end{center}

\caption{The U(1) and monopole effective string tensions
after 1 smooth on $16^4$.}
\label{fig_one}

\vspace{\figendsp}
\end{figure}

The U(1) effective string tension converges after one smoothing 
sweep to the asymptotic SU(2) value before smoothing, but with 
reduced statistical errors (Fig~\ref{fig_one}). The smoothing 
is thus only a local U(1) perturbation, in support of the 
aforementioned scenario. Interestingly, as at $\beta = 2.4$
\cite{hart98},
the monopole string tension is suppressed by ${\cal O}(30\%)$
by the single smoothing sweep. This implies that, in addition 
to magnetic monopoles, there are other objects present in 
these smoothed U(1) fields that disorder Wilson loops and 
contribute to the string tension. (Since their magnetic flux is 
conserved
these objects may be thought of as `vortices'.) 
This is not a complete surprise. If we have Abelian 
dominance (in the sense of U(1) and SU(2) plaquettes being similar)
then smoothed SU(2) fields will produce smooth U(1) fields which
cannot contain singular monopole cores. 

\begin{figure}[t]

\leavevmode
\begin{center}

\hbox{%
\hspace{\figoffset}
\epsfxsize = \figsize
\epsffile{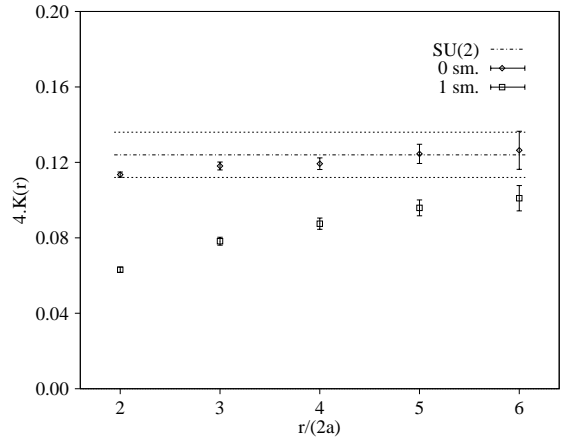}
}

\vspace{\figbackup}
\end{center}

\caption{The monopole effective string tension
after blocking on $20^4$.}
\label{fig_two}

\vspace{\figendsp}
\end{figure}

Blocking an SU(2) configuration creates a new lattice of $(L/2)^4$ 
sites with links formed by summing over staples:
$$
V_\mu(n^\prime) = D_\mu(n) + \sum_\nu
S^f_{\mu \nu}(n) + S^b_{\mu \nu}(n)
$$
where $D_\mu(n) \equiv U_\mu(n).U_\mu(n+\hat{\mu})$, 
$S^f_{\mu \nu}(n) = U_\nu(n).D_\mu(n+\hat{\nu}).
U^\dagger_\nu(n+2\hat{\mu})$,
$S^b_{\mu \nu}(n) = U^\dagger_\nu(n-\hat{\nu}).
D_\mu(n-\hat{\nu}).
U_\nu(n-\hat{\nu}+2\hat{\mu})$
and $V$ is projected back into the group.
The long range physics of such a lattice will be the same,
up to a doubling of the lattice spacing (e.g.
the asymptotic SU(2) string tension, $K$, will be four times
that of the unblocked lattice, as is plotted
here). An ensemble of (unsmoothed) SU(2) configurations,
blocked and then fixed to the MA gauge should thus show
Abelian and monopole dominance of the asymptotic string 
tension. This is so; in Fig.~\ref{fig_two} we illustrate the latter, 
and 
as before blocking $K_{\eff}(r)$ assumes its
asymptotic behaviour for small $r$. 

One smoothing sweep is sufficiently localised
that the difference between the smoothed and
unsmoothed configuration should be greatly reduced after
blocking. We thus expect at least partial
restoration of the monopole dominance that was lost under
smoothing. In Fig.~\ref{fig_two} we compare the smoothed
and unsmoothed case for a blocked $20^4$ lattice.
Although there is not a clear plateau in the effective string
tension from monopoles after blocking, in comparison to
Fig.~\ref{fig_one} the loss of monopole dominance is much
reduced.

\begin{figure}[t]

\leavevmode
\begin{center}

\hbox{%
\hspace{\figoffset}
\epsfxsize = \figsize
\epsffile{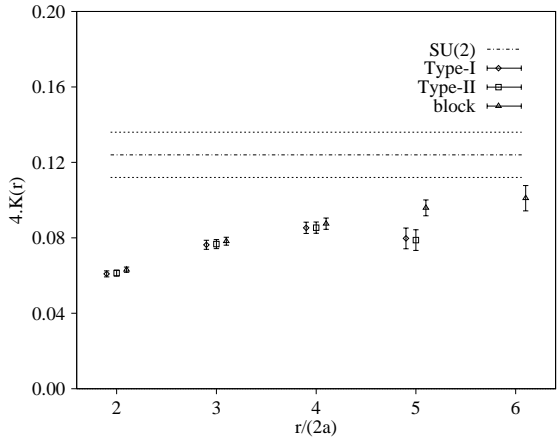}
}

\vspace{\figbackup}
\end{center}

\caption{The monopole string tensions
after 1 smooth (extended on $16^4$, blocking on $20^4$).}
\label{fig_three}

\vspace{\figendsp}
\end{figure}

\begin{figure}[t]

\leavevmode
\begin{center}

\hbox{%
\hspace{\figoffset}
\epsfxsize = \figsize
\epsffile{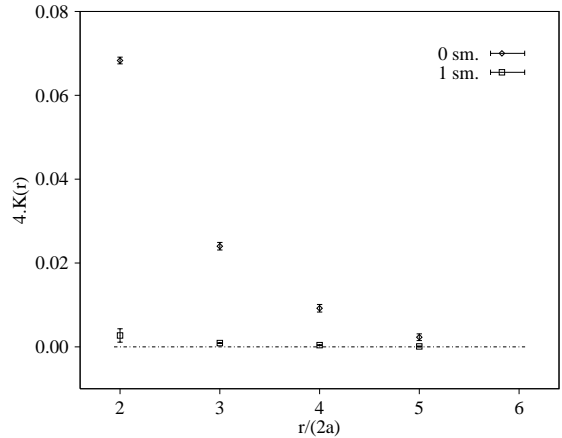}
}

\vspace{\figbackup}
\end{center}

\caption{The effective string tension for the
extended monopole difference gas on $16^4$.}
\label{fig_four}

\vspace{\figendsp}
\end{figure}

We investigated whether the effect of blocking the SU(2) links 
could be obtained by working with `extended' monopoles
\cite{ivanenko90}.
The U(1) plaquette angles are  
concentrated around multiples of $2\pi$ and interpreted as Dirac
strings. The net number of these entering an
elementary cube gives its monopole charge. Blocking these charges yields 
`Type-II' extended monopoles;
we sum the elementary charges in a $2^3$ block. Whilst this
eliminates some magnetic dipoles, it
is clear that there can be no new physics and we employ this as a control.
`Type-I' extended monopoles are found by applying the
above definition to $2\times2$ Wilson loops, and the
monopole charges assigned to $2^3$ blocks correspond to the elementary
monopoles seen after a (staple-less) blocking of the 
U(1) fields.
If using Type-I monopoles is to reproduce blocking the SU(2) links, 
Type-I monopoles should give a higher string tension than Type-II.
They are however very alike (Fig.~\ref{fig_three}). Although similar 
to that of the blocked lattices at small distances,
ultimately both Types-I and II appear to form a plateau well below the
SU(2) string tension.
In Fig.~\ref{fig_four}, we go further to compare the microscopic 
properties
of Types-I and II by locally forming a `difference gas' on each 
configuration. 
Before smoothing, there are some differences between the definitions,
but this is only short ranged. After smoothing, when we would be most
interested in Type-I identifying new magnetic
excitations, we see that they are identical at all length scales.

In conclusion, we have applied smoothing to SU(2) gauge 
configurations and found no reduction in the U(1) string 
tension after fixing to the MA gauge. This is consistent with 
the scenario in which the MA gauge isolates the long range \dof 
in the SU(2) theory. The monopole string tension, however, has 
a marked decline, ${\cal O}(30\%)$ at $\beta = 2.5$. This suggests 
the presence of additional objects in the U(1) fields capable of 
producing confinement, and also their increased r\^ole in the 
smoothed fields. These did not resemble extended monopoles, 
which failed to restore the monopole string tension. Applying a 
blocking transformation to the smoothed SU(2) fields prior to gauge 
fixing led to an approximate re--emergence of the monopole dominance. 
Previous studies here indicate a sensitivity to the smoothing method; 
Metropolis cooling of SU(2) is mild enough not to destroy the monopole 
string tension
\cite{wensley96},
but a renormalisation group based smoothing algorithm reduces even 
the U(1) string tension
\cite{kovacs97}.
There is here a complex interplay between elementary monopoles 
and other U(1) objects under smoothing, and this and the effects 
of blocking merit further study.

\newpage
\vspace{5ex}
\noindent
{\bf Acknowledgments.}

\vspace{2ex}
\noindent
The work of A.H. was supported in part by United States 
Department of Energy grant DE-FG05-91ER40167. J.D.S. was 
supported in part by the National
Science Foundation under Grant No. NSF PHY 94-12556.
M.T. was supported by United Kingdom PPARC grants GR/K55752 
and GR/K95338.

\end{document}